\documentstyle[twocolumn,aps,prl]{revtex} 
\input epsf

\newcommand{\Bbb}{I\!\!}

\begin{document}
\draft
\title{Surface critical behavior of binary alloys and antiferromagnets:
dependence of the universality class on surface orientation}
\author{Anja Drewitz$^1$, Reinhard Leidl$^1$,
Theodore\ W. Burkhardt$^{1,2}$, and H.~W.~Diehl$^1$}
\address{$^1$Fachbereich Physik, Universit\"at-Gesamthochschule Essen,
D-45117 Essen, Federal Republic of Germany}
\address{$^2$Department of Physics, Temple University, Philadelphia,
Pennsylvania 19122, USA\cite{permad}}
%\date{\today}
\date{November 20, 1996}
\maketitle
\begin{abstract}
The surface critical behavior of semi-infinite
(a) binary alloys with a continuous order-disorder transition and
(b) Ising antiferromagnets in the presence of a magnetic field is considered.
In contrast to ferromagnets, the surface universality class of these systems
depends on the orientation of the surface with respect to the crystal axes.
There is ordinary and extraordinary surface critical behavior
for orientations that preserve and break the two-sublattice symmetry,
respectively. This is confirmed by transfer-matrix calculations
for the two-dimensional antiferromagnet and other evidence.
\end{abstract}

\narrowtext
A cornerstone of the modern theory
of critical phenomena is the idea of distinct
universality classes\cite{MEF}.
The bulk universality class to which a particular system belongs
is generally determined by a few basic properties, such as
the spatial dimension and the symmetries of the order-parameter.
In the renormalization-group picture\cite{MEF} each universality class
corresponds to the basin of attraction of a fixed point in a many-dimensional
space of Hamiltonians.

Over the past two decades convincing evidence has emerged
for similar universality classes in the surface critical behavior
of semi-infinite systems close to the bulk critical point
\cite{HWD,Binder,Cardy}. The surface universality class of
a particular system is
determined by (i) the bulk universality class and
(ii) additional relevant surface properties.
In a ferromagnet in zero magnetic field, for example,
the strength of the spin couplings near
the surface is one of these additional properties. For
subcritically, critically, or supercritically enhanced
surface interactions, the surface critical behavior is ``ordinary'',
``special'', or ``extraordinary'', respectively. Another relevant surface
property is a surface ordering field. The surface critical behavior of a
ferromagnet with zero bulk magnetic field and
nonzero surface field belongs to
the ``normal'' universality class\cite{HWD-PRB94},
independent of the strength of the
surface couplings. In both the extraordinary and normal cases there is
long-range order near the surface above the bulk critical temperature, and
both transitions have the same universal properties
\cite{extra}. Following common
usage we refer to the joint universality class as extraordinary.
Note that the Ising model with $d\leq 2$ does
not exhibit a ``true'' extraordinary
transition (except for infinite surface couplings), because
of the low boundary dimension, but there is a normal transition.

Field-theoretic studies of Ising surface critical behavior \cite{HWD}
begin with the one-component $\phi^4$ model with Hamiltonian
\begin{eqnarray}\label{Ham}
{\cal H}&=&\int_{{\Bbb{R}}^d_+} 
\Big[\frac{1}{2} (\nabla \phi )^2
+\frac{\tau_0}{2}\phi^2+\frac{u_0}{4!}\phi^4\Big]
\nonumber\\&&\mbox{}
+\int_{\cal S}\left(\frac{c_0}{2}\phi^2
-h_1\phi\right)
\end{eqnarray}
defined on the half space
${\Bbb{R}}^d_+ \equiv\{(\bbox{x}_{\|},y) \in {\Bbb{R}}^d \mid y\ge 0\}$
with boundary plane ${\cal S}$ at $y=0$.
In this model there is ordinary, special, and extraordinary,
surface critical behavior at bulk criticality
for $h_1=0$ and $c_0>c_{\text{sp}}$, $c_0=c_{\text{sp}}$,
and $c_0<c_{\text{sp}}$ respectively. For
$h_1\neq 0$ and arbitrary $c_0$ the surface critical behavior is normal
\cite{Rit}.

In this Letter the surface critical behavior of
(a) binary alloys with a continuous order-disorder transition and of
(b) Ising antiferromagnets in the presence of a  magnetic field $H$
is considered. Our main result is that these systems differ from
ferromagnets in an important respect.
The surface universality class depends on {\it the
orientation of the surface plane
with respect to the crystal axes.}

That the orientation affects the surface
behavior of systems (a), (b) has been recognized
by Schmid\cite{Schmid}, who carried out Monte Carlo
simulations and mean-field calculations
for a lattice model of the A2-B2 order-disorder transition
\cite{BuchDosch} in FeAl, equivalent to an Ising antiferromagnet
with a free surface on a bcc lattice.
According to Ref. \onlinecite{Schmid}, for nonideal bulk
stoichiometry (nonzero magnetic field in the equivalent Ising antiferromagnet)
the order-parameter symmetry
is broken by the (100) surface orientation, and long-range order persists
near the surface above the bulk critical temperature $T_c$.
For the (100) surface and ideal stoichiometry or the (110) surface and arbitrary
stoichiometry, the symmetry is  not broken, and both the
bulk and surface order vanish for $T\ge T_c$.
We agree with these results but question the conclusion that
the surface critical behavior at $T_c$ is
ordinary  in all of the
above cases. We find that the surface universality class
depends on the orientation of the surface, with
extraordinary surface critical behavior for orientations that break
the symmetry of the A and B sublattices
(consistent with the
renormalization-group relevance of
surface ordering fields) and ordinary surface critical
behavior for symmetry-preserving orientations.
Transfer-matrix calculations in $d=2$
described below confirm these predictions. The predictions are also
consistent with an effective field-theoretic Hamiltonian derived
from mean-field theory, as outlined below.

As in \cite{Schmid}, we consider a simple lattice-gas model in which
each site of a bcc lattice is occupied by a particle of type 1 or 2.
Nearest-neighbor pairs contribute $V_{11}$, $V_{22}$, or $V_{12}$,
depending on the species involved, to the total energy.
In the grand canonical ensemble with chemical potentials $\mu_1, \mu_2$
the lattice gas is equivalent to an Ising model with
Boltzmann factor $\exp (-{\cal H}_{\text{lat}})$, where
\begin{equation}\label{latHam}
{\cal H}_{\text{lat}} = K\sum_{\langle \bbox{i},\bbox{j}\rangle}
s_{\bbox{i}} s_{\bbox{j}}
- H\sum_{{\rm all}\ \bbox{i}} s_{\bbox{i}}
 - H_1\sum_{\bbox{i}\in{\cal S}}s_{\bbox{i}}\;.
\end{equation}
The bulk and excess surface magnetic fields are given by
$H={1\over 2}[(\mu_1-\mu_2)-{1\over 2}
\zeta\thinspace (V_{11}-V_{22})](k_BT)^{-1}$
and $H_1={1\over 4}(\zeta-\zeta_1)(V_{11}-V_{22})](k_BT)^{-1}$,
where $\zeta$ and $\zeta_1$ are the coordination numbers of
interior and surface sites, respectively. The nearest-neighbor coupling
$K={1\over 4}(V_{11}+V_{22}-2V_{12})(k_BT)^{-1}$ is assumed to be positive,
corresponding to an antiferromagnet, and is the same
for surface and interior spins. For $\vert H\vert K^{-1}<\zeta$,
$\vert H+H_1\vert K^{-1}<\zeta_1$, the system is antiferromagnetically
ordered at $T=0$. The two-dimensional analog of the model is shown
in Fig.~1, with the A and B sublattices indicated by
filled and empty points.

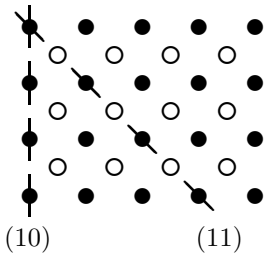
\begin{figure}
\begin{center}
\unitlength0.75mm
\begin{picture}(45,45)\thicklines
\multiput(2.5,8)(0,10){4}{\circle*{3}}
\multiput(7.5,13)(0,10){3}{\circle{3}}
\multiput(12.5,8)(0,10){4}{\circle*{3}}
\multiput(17.5,13)(0,10){3}{\circle{3}}
\multiput(22.5,8)(0,10){4}{\circle*{3}}
\multiput(27.5,13)(0,10){3}{\circle{3}}
\multiput(32.5,8)(0,10){4}{\circle*{3}}
\multiput(37.5,13)(0,10){3}{\circle{3}}
\multiput(42.5,8)(0,10){4}{\circle*{3}}
\multiput(2.5,4.25)(0,10){4}{\line(0,1){7.5}}
\multiput(0,40.5)(10,-10){4}{\line(1,-1){5}}
\put(-2,-1){(10)}\put(32,-1){(11)}
\end{picture}
\end{center}
\caption{Symmetry breaking (10) surface (vertical broken line)
and symmetry preserving (11) surface (diagonal broken line).}
\end{figure}

The standard definition of the bulk order
parameter of the antiferromagnet in a magnetic field is the difference
of the sublattice magnetizations. The conjugate ordering field
is a staggered magnetic field. The bulk field $H$
in (\ref{latHam}) is an example
of a nonordering field. Increasing $H$ lowers the critical temperature,
at least for weak fields, but does not alter the bulk universality class.
As can be seen by redefining the spins on one sublattice with a minus sign,
the Ising antiferromagnet is exactly equivalent to an Ising
ferromagnet with a nonordering staggered field, irrelevant in the bulk.
Thus the same universality classes as in the ferromagnetic
Ising model are expected.

The relevance of the surface orientation can be understood from
Fig.~1. Suppose that there are $N\to\infty$ layers
of spins, with free surfaces at the first
and $N$th layers and periodic boundary conditions in the
other direction. For the (11) surface orientation adjacent spins in
any layer belong to different sublattices.
Both the bulk and surface fields $H$ and $H_1$ in Eq.\ (\ref{latHam})
are nonordering. The Hamiltonian is invariant
under a translation of all the spins parallel to the surface by one lattice
constant, i.e., under interchange of the A and B lattices. Thus the (11)
orientation respects the symmetry of the two sublattices.
The order-parameter profile $\phi_n$,
defined as the difference of the sublattice magnetizations in the $n$th layer,
vanishes for $T>T_c$ due to symmetry of the sublattices. For $T<T_c$
there is ordering at the surface driven by the bulk order. Thus ordinary
surface critical behavior is expected.

A (10) surface breaks the A-B symmetry of the antiferromagnet,
since all the surface spins belong to a particular sublattice.
The order-parameter profile is symmetric and antisymmetric
about the midpoint of a strip with $N$ layers for $N$ odd and even,
respectively (see Fig.~3 and Fig.~4 below).
This can be understood in terms of the equivalent ferromagnet.
The bulk staggered field is
constant within a layer but alternates in sign from layer to layer.
The fields of magnitude $|H+H_1|$ on the two surfaces, which are parallel
for $N$ odd and antiparallel for $N$ even, are ordering fields for this
orientation. Under the renormalization group
the bulk staggered field is driven to zero, but surface fields
with $++$ or $+-$ orientations survive.
For both the $++$ and $+-$ orientation extraordinary critical behavior
is expected in the limit $N\to\infty$, due to the field-induced order
at the surface.
 
Applying the same reasoning to the three-dimensional analog of the above
system (bcc lattice, $N\to\infty$ layers, two free surfaces,
and periodic boundaries in all other directions),
we predict ordinary and extraordinary surface critical
behavior for (110) and (100) surfaces, respectively.

The above conclusions also apply if either
of the two fields $H,\ H_1$ vanishes.
When $H=H_1=0$ (ideal stoichiometry)
the magnetization vanishes on both sublattices for $T>T_c$.
Ordinary critical behavior is expected for
arbitrary surface orientations.

We have checked our predictions in $d=2$ with numerical transfer-matrix
methods. First the correlation length $\xi_N(K_c(H),H,H_1)$
parallel to (11) and (10) surfaces of the antiferromagnet (\ref{latHam})
defined on a strip of square lattice with $N$ layers of spins
was calculated. Here $K_c(H)$ is the bulk critical
coupling constant, determined as a function of $H$ with great precision
in Ref.\ \onlinecite{critline}.
Then the universality class was deduced by
comparing $\pi N^{-1}\xi_N$,
with $\xi_N$ expressed in units of the layer separation,
with the known amplitudes
\begin{equation}
{\cal A}^{ab}=\pi\lim_{N\to \infty} N^{-1}\xi_N^{ab}
\end{equation}
for ferromagnetic Ising strips with
boundary conditions $a,b$.
The ${\cal A}^{ab}$, which are
clearly scale invariant, are universal quantities that depend on
the surface universality class. For ferromagnetic Ising strips
with ordinary (zero-field) boundary conditions and with parallel
and antiparallel surface fields
\begin{equation}\label{amp}
{\cal A}^{\text{ord,ord}}=2\ ,\qquad
{\cal A}^{++}={1\over 2}\ ,\qquad
{\cal A}^{+-}=1\ .
\end{equation}

The first two amplitudes in Eq.\ (\ref{amp}) follow from
the exact \cite{2dexact} results $\eta_\|^{\text{ord}}=1$,
$\eta_\|^{\text{ex}}=4$ for the semi-infinite two-dimensional
Ising model and Cardy's \cite{Cardy} formula
${\cal A}^{aa}=2(\eta_\|^a)^{-1}$
for homogeneous critical systems with the same boundary condition $a$ on
both edges. This relation is a consequence of conformal
invariance and the conformal mapping of the half plane with boundary condition
$a$ onto the strip. Here the surface critical
exponent $\eta_\|$ is defined by the $r^{-\eta_\|}$ decay of
order-parameter correlations parallel to the boundary of the half space.
The entry ${\cal A}^{+-}=1$
in Eq.\ (\ref{amp}) is derived in Ref.\ \onlinecite{Cardy3}.
It also follows from the exact spin-spin correlation
function \cite{BX} for $+-$ boundary conditions.

We have numerically analyzed the transfer matrices
of rectangular slices of the strip
with dimensions $(N-1)\times\tau$, where  $\tau=1$ for (11) edges and $\tau=2$
for (10) edges. The correlation length $\xi_N$, expressed, like
the width $N-1$ of the strip and $\tau$,
in units of the layer separation, was determined from the
largest and next largest eigenvalues of the transfer matrix
$\lambda_N^{(1)}$ and $\lambda_N^{(2)}$ using the relation
\begin{equation}
\xi_N^{-1}=\tau^{-1}\ln\big|\lambda_N^{(1)}/\lambda_N^{(2)}\big|\;.
\end{equation}
Writing the transfer matrix as a product of sparse matrices
\cite{Bloete}, we obtained results up to $N=19$ layers for the (11)
orientation and $N=37$ for the (10) orientation.
For more details see Ref.\ \onlinecite{AD-DA}.

\begin{figure}
\centerline{\epsfbox{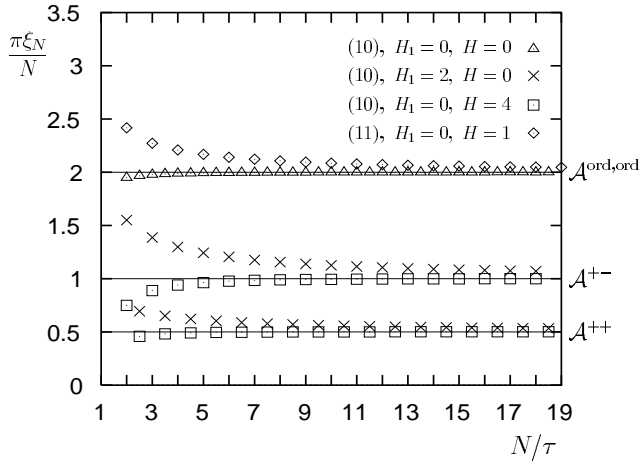}}
\smallskip
%%%% use \vspace if figure is not included by \epsfbox
%%%% \vspace*{61mm}
\caption{Transfer-matrix results for strips of $N$ layers
with (10) and (11) surfaces. For the (10) and (11) orientations,
$\tau=2$ and $1$, respectively. The middle two and lower two
curves show results for even and odd $N$, respectively.}
\end{figure}

Representative numerical data are shown in Fig.~2.
The quantity $\pi N^{-1}\xi_N$ extrapolates convincingly to
the amplitudes $2$, ${1\over 2}$, and $1$ in Eq.\  (\ref{amp}),
in agreement with the surface universality classes predicted above.

The ordinary surface critical behavior expected in system (\ref{latHam})
for the (11) orientation and arbitrary values of $H$, $H_1$,
and for the (10) orientation in the case $H= H_1 =0$
(ideal stoichiometry) is confirmed by the transfer-matrix results.
The corresponding data in Fig.~2 agree well with
${\cal A}^{\text{ord,ord}}=2$.

We predict extraordinary surface critical behavior for the (10) orientation
with nonvanishing $H$, $H_1$, due to effective ordering surface fields,
with $++$ and $+-$ orientations for
antiferromagnets with odd and even $N$, respectively.
This is also confirmed by the transfer-matrix results. The corresponding
data in Fig.~2 are in excellent agreement with
${\cal A}^{++}={1\over 2}$ for $N$ odd
and ${\cal A}^{+-}=1$ for $N$ even.
The magnetization and order-parameter profiles in strips of $37$ and
$36$ layers with (10) edges, calculated from the eigenvector
with the largest eigenvalue $\lambda_N^{(1)}$,
are shown in Fig.~3 and Fig.~4, respectively.

\begin{figure}
\centerline{\epsfbox{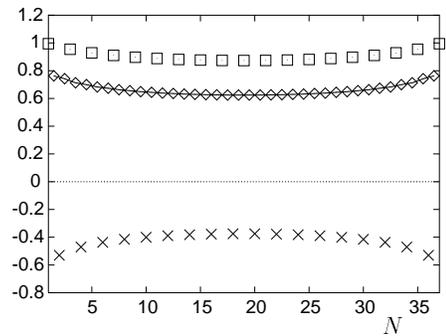}}
%%%% use \vspace if figure is not included by \epsfbox
%%%% \vspace*{45mm}
\smallskip
\caption{Magnetization profile of a strip of $N=37$ layers
with (10) surfaces and $H_1=0$, $H=4$, $K=K_c(H)$.
Sublattices A and B are indicated
by $\Box$ and $\times$, respectively.
The order parameter ($\Diamond$) is defined as half
the difference of the A and B sublattice magnetizations in adjacent layers.}
\end{figure}

\begin{figure}
\centerline{\epsfbox{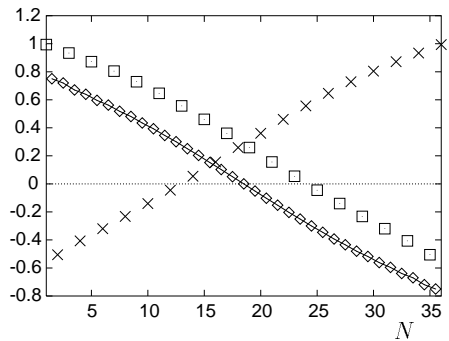}}
%%%% use \vspace if figure is not included by \epsfbox
%%%% \vspace*{45mm}
\smallskip
\caption{Same as Fig.~3, but with $N=36$.
Since the surface spins on opposite edges belong to different sublattices,
the order-parameter profile is antisymmetric.}
\end{figure}

Finally we consider the implications of mean-field theory for
the surface universality class. Mean-field theories are, of course,
of interest as tractable first approximations. More importantly,
one can often infer the appropriate continuum Hamiltonian
${\cal H}[\phi]$ for renormalization-group analysis from
mean-field theory. The standard procedure is to begin with
mean-field difference equations on a lattice and taking the continuum
limit. The appropriate ${\cal H}[\phi]$ is minimized by the resulting
differential equations and boundary conditions, which correspond
to the Landau equations $\delta{\cal H}/\delta \phi =0$.

As in Ref.\ \onlinecite{Schmid}, we have carried out mean-field
studies, but with some differences in approach and conclusions.
Here we give a brief summary. The details will be
published separately\cite{Leidl-MFT,Leidl-LandauTheory}.

It turns out to be very useful to interpret the rather complex mean-field
difference equations as a nonlinear recursive map, an approach
pioneered by Pandit and Wortis \cite{PW}. For the various surface
orientations considered above, the mapping corresponds to
discrete Hamiltonian dynamics. The general properties of Hamiltonian
flows have been studied extensively\cite{nldynamics}. The
nonvanishing order parameter profile
for $T\ge T_c$ in the case of a symmetry breaking
surface orientation can be understood in this context.

Our mean-field equations also provide a convenient
starting point for the continuum approximation and, in our opinion, avoid
some problematic aspects of earlier work \cite{Schmid}.
We are led, apart from irrelevant terms, to the familiar $\phi^4$
Hamiltonian (\ref{Ham}) for semi-infinite systems, with a nonzero surface
field $h_1$ in the case of
symmetry-breaking surface orientations and zero surface field
otherwise. The question of surface fields in the
continuum Hamiltonian is conveniently analyzed in the
framework of  Landau-like symmetry arguments \cite{LL-Vol-V} and
the ``method of concentration waves'' \cite{Khach}. As explained in
Ref.\ \onlinecite{Leidl-LandauTheory}, the bulk
amplitudes of all concentration waves that do not share
the symmetry of the order parameter vanish in the high-temperature
phase. In the presence of a planar boundary translational
invariance is lost, and there are some nonzero amplitudes.
The effective surface field for symmetry-breaking
surface orientations arises from a coupling of the order parameter
to nonvanishing concentration waves.

In summary, we have shown that the surface universality class of
semi-infinite binary alloys and antiferromagnets depends on the
surface orientation, with ordinary and extraordinary behavior for
symmetry-preserving and symmetry-breaking orientations, respectively.
The underlying mechanism is rather general, and Ising spins are not
essential. We expect the orientational dependence to be a common
feature of semi-infinite systems with (i) nonordering fields and
(ii) second-order phase transitions in which the symmetry
of two or more spatially distinct sublattices
is spontaneously broken below $T_c$ in the bulk.

We thank Henk W.\ J.\ Bl\"ote for useful correspondence.
T.~W.~B. appreciates the support of the Alexander von Humboldt-Stiftung.
This work was also supported by the Deutsche Forschungsgemeinschaft
through Son\-der\-for\-schungs\-be\-reich 237.

\end{document}